# Individual nanoantennas empowered by bound states in the continuum for nonlinear photonics

Kirill Koshelev[1,2], Sergey Kruk[1], Elizaveta Melik-Gaykazyan[1,3], Jae-Hyuck Choi[4],

Andrey Bogdanov[2], Hong-Gyu Park[4], and Yuri Kivshar[1,2]

[1]Nonlinear Physics Center, Australian National University, Canberra ACT 2601, Australia
[2]ITMO University, St Petersburg 197101, Russia
[3]Faculty of Physics, Lomonosov Moscow State University, Moscow 119991, Russia
[4]Department of Physics, Korea University, Seoul 02841, Republic of Korea

**Bound states in the continuum (BICs) represent localized modes with energies embedded in the continuous spectrum of radiating waves. BICs were discovered initially as a mathematical curiosity in quantum mechanics[1], and more recently were employed in photonics[2,3]. Pure mathematical bound states have infinitely-large quality factors (Q factors) and zero resonant linewidth. In optics, BICs are physically limited by a finite size, material absorption, structural disorder, and surface scattering[4], and they manifest themselves as the resonant states with large Q factors, also known as *supercavity modes* or *quasi-BICs*. Optical BIC resonances have been demonstrated only in extended 2D and 1D systems[5-9] and have been employed for distinct applications including lasing[10,11] and sensing[12]. Optical quasi-BIC modes in individual nanoresonators have been discovered recently[13,14] but they were never observed in experiment. Here, we demonstrate experimentally an isolated subwavelength nanoresonator hosting a quasi-BIC resonance. We fabricate the resonator from AlGaAs material on an engineered substrate, and couple to the quasi-BIC mode using structured light. We employ the resonator as a nonlinear nanoantenna and demonstrate record-high efficiency of second-harmonic generation. Our study brings a novel platform to resonant subwavelength photonics.**

Engineering nonradiative optical states in extended systems such as photonic crystals and metasurfaces is well-studied, especially for structures possessing certain symmetries[4-12]. For complete cancellation of total radiative losses in extended structures below the diffraction limit, it is sufficient to suppress energy leakage in the only allowed radiation direction. On the contrary, for a localized resonator any portion of trapped light is released through a continuum of radiation modes simultaneously, which leads to impossibility of nonradiative states[2] in compact geometries, unless one considers uncommon for optical frequencies materials with zero permittivity[15]. The concept of quasi-BICs provides an elegant solution for engineering high-quality individual nanoantennas operating in the subwavelength regime as it is based on suppression of the radiation loss via destructive interference between different modes. Moreover, the modes forming a quasi-BIC belong to the same resonator which allows to keep the sample footprint small giving a strong benefit over resonators relying on alternative mechanisms of localization, including whispering gallery modes and cavities in photonic-bandgap structures, where an efficient light trapping requires dozens of structural periods.

Here, we also employ the quasi-BIC resonances for the second-harmonic generation. Traditionally, nonlinear optical processes rely on the propagation of light over long distances inside nonlinear crystals or optical waveguides[16]. However, subwavelength optical resonators supporting localized modes allow realizing efficient nonlinear processes at the nanoscale. Many demonstrations with nonlinear nanoantennas employ plasmonic resonances in metallic nanoparticles[17], but efficiencies of nonlinear processes in such structures remain extremely low[18,19]. Resonant dielectric nanophotonics[20] opens a novel direction towards efficient nonlinear processes in the subwavelength regime[13,14,21]. Here, we demonstrate that our subwavelength resonator facilitating strong confinement of light increases nonlinear response by orders of magnitude offering a paradigm shift in nonlinear nanophotonics.

We design the quasi-BIC mode in a nonlinear nanoantenna of cylindrical shape made of AlGaAs (aluminum gallium arsenide) placed on a three-layer substrate ($SiO_2$/ITO/$SiO_2$) (Fig. 1a). The physics of quasi-BIC relies on destructive interference of two leaky modes with similar far-field patterns. To couple incident light efficiently to the quasi-BIC resonant mode from the far-field, we employ an azimuthally polarized vector beam. The cylindrical resonator supports two families of leaky modes[13], which can be understood as radially and axially polarized modes, as depicted in Fig. 1b. The modes of the resonator inherit its structural symmetry and can be additionally classified according to their azimuthal order. We select a pair of radial and axial modes with the zero azimuthal index which both demonstrate dipolar magnetic response in the far-field zone (see Supplementary Information). Decreasing the difference between their frequencies via tuning of the disk diameter we achieve the strong coupling regime which produces the characteristic avoided resonance crossing of frequency curves and modification of mode radiative Q factors (see Figs. 1c,d). As shown in Fig. 1d, the mode with the highest Q factor is realized at a specific diameter in the vicinity of the avoided resonance crossing. In our case, the quasi-BIC appears for the diameter of 930 nm and a height of 635 nm.

We notice that, when a nanoresonator is placed on a substrate, the optical contrast between the resonator and its environment gets reduced, typically negatively affecting the mode Q factor[22]. To compensate the decrease of the Q factor induced by energy leakage into the substrate, we engineer the substrate with an additional layer of ITO (indium tin oxide) exhibiting epsilon-near-zero transition acting as a conductor above 1200 nm wavelength (e.g. at the quasi-BIC wavelength), and as an insulator below this wavelength. The ITO layer is separated from the resonator by a $SiO_2$ spacer. Thickness of the $SiO_2$ spacer layer allows to control the phase of reflection, also enhancing the destructive interference of the two leaky modes in the far-field and thus increasing the Q factor. This can be understood as an interaction of the resonator with its "image" in the substrate[23,24]. As shown in Fig. 1e, the Q factor of the quasi-BIC mode oscillates with the spacer thickness approaching a constant value corresponding to a bulk $SiO_2$ substrate. The Q factor peaks for spacer thicknesses between 300 and 400 nm with the maximal value of 235 predicted theoretically. This value is more than five times higher compared to a nanoantenna placed on a bare glass and two times higher compared to a free-standing nanoantenna in air.

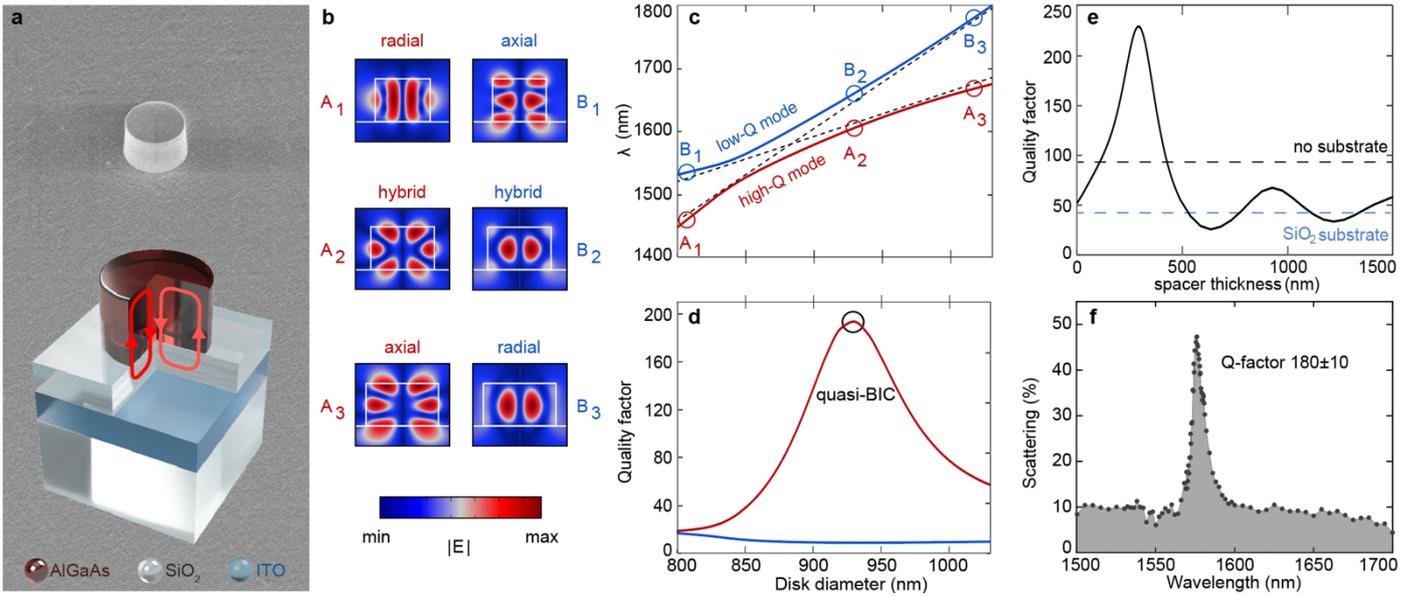

**Figure 1. Optical quasi-BIC mode in an isolated subwavelength nanoparticle.** (a) Scanning electron microscope image (top) and schematic (bottom) of an isolated subwavelength nanoantenna supporting high-Q modes. (b) Calculated near-field patterns of the two leaky modes hosted by the resonator and (c) their wavelengths vs. resonator diameter. Strong coupling of radial and axial resonator modes with similar far-field patterns results in the emergence of a quasi-BIC mode. (d) Calculated Q factor for the low-Q and high-Q modes vs. resonator diameter. Calculations for (c) and (d) are done for a 350 nm $SiO_2$ spacer. (e) Calculated Q factor of the quasi-BIC vs. $SiO_2$ spacer thickness. The dashed lines show the values of the Q factor for a disk in air and a disk on a bulk $SiO_2$ substrate. (f) Measured scattering spectrum and retrieved Q factor of the observed resonance.

We fabricate such nanoantennas from epitaxially grown AlGaAs (crystal axes orientation [100], 20% Al) by means of electron-beam lithography and dry etching process (details of the fabrication process available in Methods). For the substrate, we use commercial films of 300 nm ITO on $SiO_2$ on top of which we deposit a 350 nm $SiO_2$ spacer. We finally transfer AlGaAs nanoparticles to the $SiO_2$/ITO/$SiO_2$ substrate. For the fabricated structure, we perform linear measurements using azimuthally polarized excitation and an objective lens (×100, 0.7NA) to focus the pump beam and to collect the backward-scattered signal. Figure 1f shows the difference between the measured backward scattering of the bare substrate and the fabricated sample normalized to the substrate reflectivity. The experimental value of the Q factor of about 180 is extracted from the measured spectrum (see Methods).

We employ the high-Q resonance of the nanoparticle to achieve a sharp enhancement of the second-harmonic generation (SHG) conceptually depicted in Fig. 2a. The AlGaAs material is known for its superior second-order nonlinear properties. In conventional nonlinear optics of long, macroscopic structures, it is necessary to compensate for optical dispersion, which results in different phase velocities for light of different frequencies[25]. In contrast, at the subwavelength scale and beyond the applicability of phase-matching, both geometric and material resonances drive the optical response.

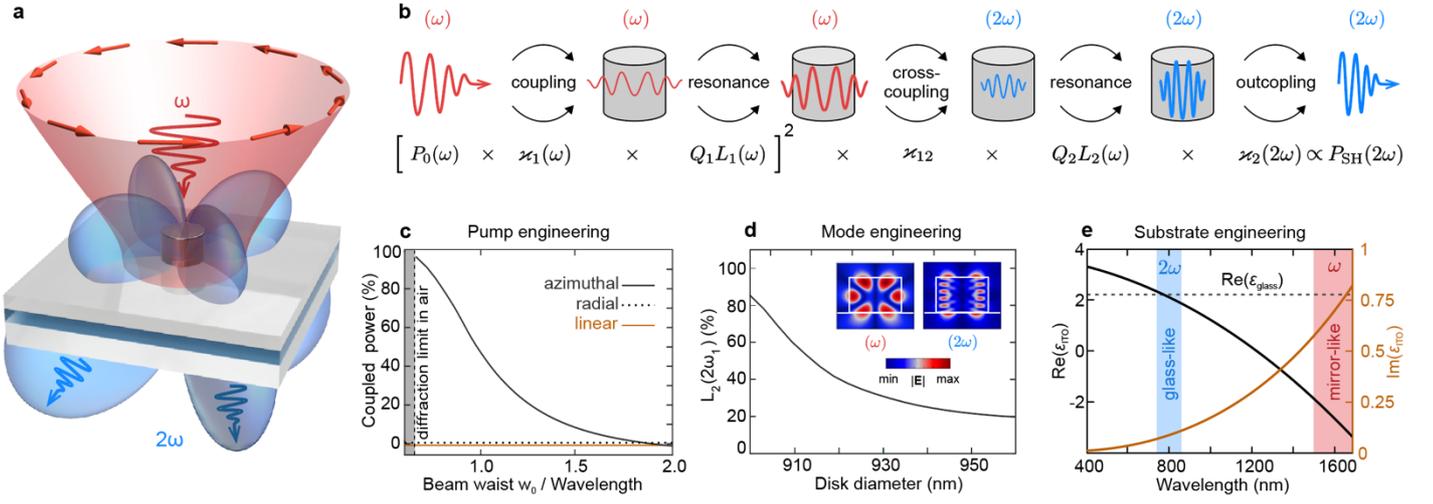

**Figure 2. Enhancement of nonlinear effects in quasi-BIC modes.** (a) Illustration of the SHG in the quasi-BIC resonator on a multilayered substrate under azimuthally polarized vector beam excitation. (b) Schematic diagram of SHG in nanoscale resonators. Each term of the formula describes one stage of the process. (c) Percent of the pump power coupled to the quasi-BIC depending on the ratio between the beam waist and the pump wavelength. The calculation is done for a free-standing nanoantenna in air. Different colors show azimuthally, radially, and linearly polarized cylindrical pump beams. (d) Spectral overlap between the quasi-BIC at the pump frequency and the magnetic quadrupole mode at the second-harmonic frequency $L_2(2\omega_1)$ vs. the disk diameter. The inset shows the near-field profiles of the modes at the pump frequency and at the second-harmonic frequency. (e) Experimental ellipsometry data for the permittivity of the ITO layer. Experimental ranges of the pump power and collected SH power are marked with the red and blue areas, respectively.

To understand how to achieve high levels of the nonlinear frequency conversion, we develop the detailed step-by-step theory of SHG at the nanoscale. We employ the eigenmode expansion method for non-Hermitian electromagnetic systems proved to be a powerful tool to understand the peculiarities of light scattering[26]. We extend this approach to describe the resonant response in the nonlinear regime (see Supplementary Information).

The designed subwavelength nanoantenna supports a quasi-BIC mode with frequency $\omega_1$ and inverse radiation lifetime $\gamma_1$ and a magnetic quadrupole Mie resonance with frequency $\omega_2$ and inverse radiation lifetime $\gamma_2$ in the vicinity of the SHG wavelength (see Supplementary Information). To find the resonance amplitudes we expand the resonator's Green's function over the eigenmodes both at the fundamental and SHG frequencies and derive the expression for the total second-harmonic power radiated by the subwavelength disk (see details in Supplementary Information)

$$P_{\text{SH}}(2\omega) = \alpha \, \varkappa_2 Q_2 \, L_2(2\omega) \, \varkappa_{12} \, [Q_1 \, L_1(\omega) \, \varkappa_1(\omega) \, P_0(\omega)]^2. \qquad (1)$$

This expression allows a detailed step-by-step treatment of the SHG process, whose schematic diagram is shown in Fig. 2b. The incident power $P_0$ is coupled to the quasi-BIC with efficiency governed by the spatial overlap $\varkappa_1$ between the pump and the mode. The coupled power is resonantly enhanced depending on the mode Q factor $Q_1$ and damped by the spectral overlap factor $L_1(\omega) = \gamma_1^2/[(\omega - \omega_1)^2 + \gamma_1^2]$ which tends to the unity at the resonance. The efficiency of up-conversion of the total accumulated power is determined by the cross-coupling coefficient $\varkappa_{12}$ which depends on the symmetry of the nonlinear susceptibility tensor of AlGaAs and the spatial overlap between the generated nonlinear polarization current and the quadrupole Mie resonance in the vicinity of the SHG wavelength (see Supplementary Information). The converted second-harmonic power is increased by a high Q factor of the second-harmonic mode but at the same time is decreased due to the spectral mismatch with the quasi-BIC

$$L_2(2\omega_1) = \gamma_2^2 / [(2\omega_1 - \omega_2)^2 + \gamma_2^2]. \qquad (2)$$

The outcoupling factor $\varkappa_2$ which tends to the unity for the frequencies in the vicinity of $\omega_2$ determines the fraction of the radiated second-harmonic power. The exact expressions for coupling coefficients $\varkappa_1, \varkappa_{12}, \varkappa_2$ and the constant $\alpha$ are given in Supplementary Information.

The developed approach allows to specify the optimal conditions to maximize the SHG efficiency from an individual AlGaAs nanoantenna. First, the spatial profile of the pump must be structured to match the distribution of the excited mode. To provide the perfect spatial matching, we use the cylindrical vector beam with azimuthal polarization resembling the far-field pattern of the quasi-BIC. We estimate $\varkappa_1$ as 33% for the experimental conditions using a model of a free-standing nanoantenna in air, as demonstrated in Fig. 2c. Next, the optimal structure must be resonant simultaneously at pump and second-harmonic wavelengths and both resonances must be of a high Q factor. For the designed subwavelength disk (Al 20%) with the diameter of 930 nm, the factor of spectral overlap reaches 30%, as shown in Fig. 2d. At the SHG wavelength, the nanoparticle hosts a quadrupolar mode with a Q factor $Q_2 \approx 65$ (see Supplementary Information). Finally, the collection efficiency must be increased, which can be done by optimizing the substrate properties. The ITO film of the structured substrate, while acting as a mirror at the pump wavelength, is transparent at the SHG wavelength with material properties similar to glass as shown in Fig. 2e. The epsilon-near-zero transition occurring in the ITO makes it effectively "invisible" to the second-harmonic radiation allowing it to propagate in both forward and backward directions.

To do systematic experimental analysis of the SHG enhancement in quasi-BIC resonators, we fabricate a set of individual AlGaAs disks with a varying diameter from 890 to 980 nm. We excite the resonators in the wavelength range from 1500 to 1700 nm using laser pulses with 2 ps duration (see experimental details in Methods). The nanoparticle is positioned between two confocal objective lenses: the first lens (×100, 0.7NA) focuses the pump beam and collects backward-generated second harmonic. It is achromatic over the spectral range of both the pump and the second harmonic. The other lens (×100, 0.9NA) collects the second harmonic in forward direction. We employ azimuthally, radially, and linearly polarized pump to experimentally verify the selective spatial coupling to the mode at the pump wavelength.

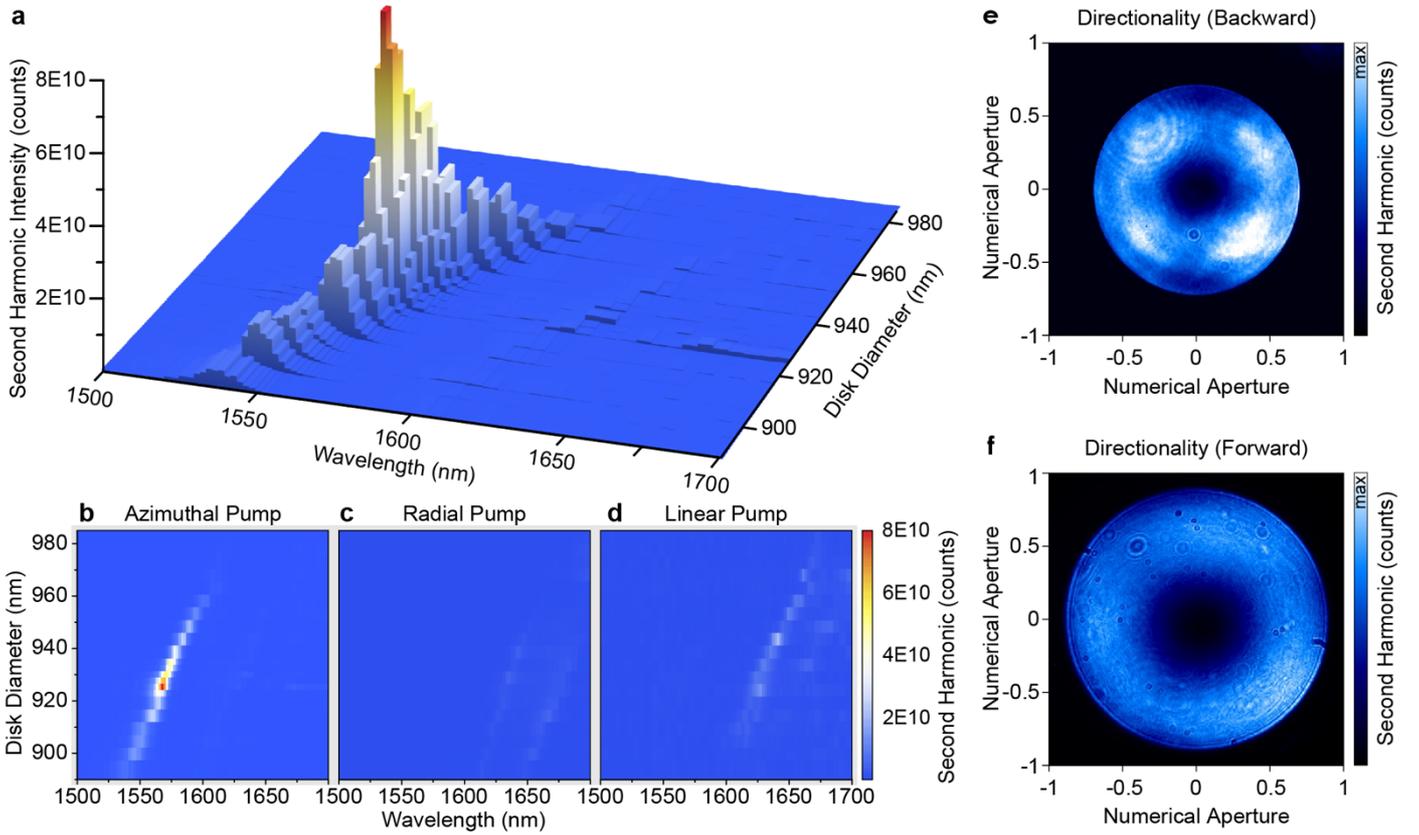

**Figure 3. Experimental characterization of the SHG enhancement in subwavelength resonators.** (a) 3D map of second-harmonic power measured as a function of the pump wavelength and resonator diameter for an azimuthally polarized beam. (b-d) Top views on the maps of SHG with the (b) azimuthal, (c) radial, and (d) linear pump. (e,f) Experimentally measured directionality diagrams of the SHG for the nanoantenna with the diameter of 930 nm in the (e) backward and (f) forward direction.

Figure 3 summarizes our experimental results for the SHG enhancement in individual disks. Figures 3a-d show maps of the SHG intensity as a function of both the pump wavelength and resonator diameter for the disks pumped by the azimuthal, radial, and linearly polarized beam, respectively. For the linearly polarized excitation, the polarization angle with respect to the AlGaAs crystalline axes is chosen such that it provides the highest second-harmonic signal. The experiment reveals an exceptionally high enhancement of the nonlinear signal in the quasi-BIC regime for the azimuthally polarized pump. Figures 3e,f show experimentally measured directionality diagrams of the second-harmonic signal in both backward (0.7NA) and forward direction (0.9NA) within the numerical aperture of the objective lens. The diagram in backward direction features distinct maxima in four directions. This is qualitatively similar to the theoretical SHG directionality shown in Fig. 2a and the far-field pattern of the mode excited at the second-harmonic wavelength (see Supplementary Information).

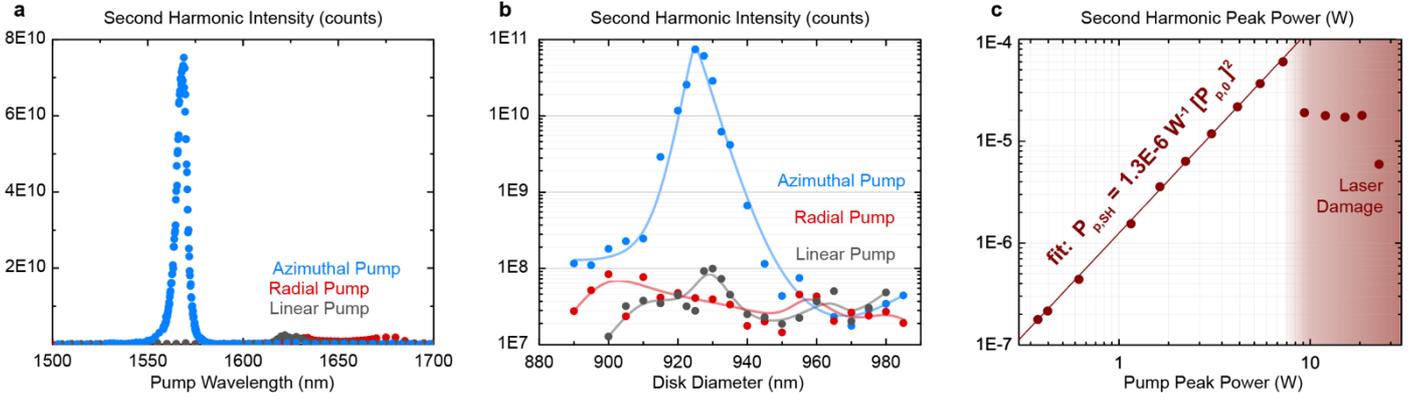

**Figure 4. Nonlinear conversion efficiency.** (a) The second-harmonic intensity as a function of the pump wavelength for the nanoantenna with the diameter of 930 nm for all three pump polarizations. (b) Measured second-harmonic intensity as a function of the nanoantenna diameter for all three pump polarizations at the wavelength 1570 nm. (c) Measured peak second-harmonic power vs. the peak pump power for the nanoantenna with the diameter of 930 nm ($\text{Log}_{10}-\text{Log}_{10}$ scale). Line shows the fit with a quadratic dependence $P_{p,\text{SH}}(P_{p,0})$ and nonlinear conversion coefficient $1.3\times10^{-6}$ W$^{-1}$.

Figures 4a,b show a wavelength cut (at the quasi-BIC diameter of 930 nm) and a size cut (at the quasi-BIC wavelength of 1570 nm) of the measured 2D maps (see Figs. 3b-d). Both plots demonstrate that the observed second-harmonic intensity for the azimuthal pump surpasses the second-harmonic intensity for the other polarizations by orders of magnitude, which confirms high spatial selectivity of the quasi-BIC. We further experimentally measure the SHG conversion efficiency by detecting the pump power incident onto the sample, and second-harmonic power captured by the two objective lenses in forward and backward directions. The resulting measurements are shown in Fig. 4c. Directly measured conversion efficiency as the ratio between the collected peak second-harmonic power $P_{p,\text{SH}}$ and the incident power $P_{p,0}$ squared is $1.3\times10^{-6}$ W$^{-1}$. The nonlinear conversion efficiency demonstrated here is more than two orders of magnitude higher than earlier results for individual nanoresonators[19,27-30] which makes it a significant step forward for the development of the efficient nonlinear optics at the nanoscale. We further evaluate the total SHG efficiency using the common approach by comparing the total radiated second-harmonic power estimated using the collection objective NA with the percent of the pump power coupled to the resonator. We take into account theoretically estimated 33% of pump coupling efficiency, and 24% of SHG collection efficiency, and obtain an estimated total conversion efficiency of $4.8\times10^{-5}$ W$^{-1}$. A comprehensive list of experimental parameters and elaborated comparison with earlier results for individual nanoantennas[19,27-30] is presented in Supplementary Information.

We finally briefly note on the SHG efficiency of a single individual resonator demonstrated here compared to the efficiencies of nonlinear metasurfaces (e.g. arrays of nanostructures extended in two dimensions). In that area, best efficiencies to date have been demonstrated by coupling nanostructured metasurfaces to quantum wells[31-33]. In these works, *linear* response of a metasurface (metallic or dielectric) provides pump coupling to a *nonlinear* quantum well. Although high nonlinear coefficients $P_{p,\text{SH}}/P_{p,0}^2$ [W$^{-1}$] have been demonstrated in such systems in the far-infrared to mid-infrared spectral ranges, the overall efficiencies in these structures $P_{p,\text{SH}}/P_{p,0}$ [%] remain limited by levels of pump power that they can sustain. Efficiencies of $2\times10^{-4}$ %,[31] $2.5\times10^{-3}$ %,[32] and $7.5\times10^{-2}$ %[33] have been reported. Also, the direct quantitative comparison of a single subwavelength resonator to a metasurface cannot be done without some ambiguity, we generally note that the efficiency demonstrated here for an individual particle is comparable to two-dimensional nonlinear metasurfaces.

In summary, we have demonstrated experimentally, for the first time to our knowledge, the existence of high-Q optical modes in individual subwavelength resonators governed by the physics of bound states in the continuum, accessed with a structured light and enhanced by an engineered substrate. We have employed these quasi-BIC resonant nanoantennas for nonlinear nanophotonics and observed the record-high second-harmonic generation at the nanoscale empowered by the enhanced field confinement associated with this novel type of high-Q resonant optical mode. Our demonstration opens novel opportunities for resonant nanoscale optics and photonics. High-Q modes supported by individual isolated nanoantennas will find further applications in subwavelength lasers, sensors and quantum sources.

**METHODS**

**Sample fabrication.** We epitaxially grow AlGaAs (20% Al) film on a buffer AlInP layer on top of a [100] GaAs substrate. The thickness of the AlGaAs film is about 635 nm. Also, we deposit a $SiO_2$ spacer on the ITO substrate by plasma-enhanced chemical vapour deposition. The thickness of the $SiO_2$ spacer is about 350 nm. The AlGaAs disk resonators are fabricated with an electron-beam lithography with chemically assisted ion-beam etching. The AlInP buffer layer is wet etched by diluted HCl solution. Then, the AlGaAs disks are attached to a polypropylene carbonate (PPC)-coated polydimethylsiloxane (PDMS) stamp. The AlGaAs resonators are transferred to the $SiO_2$/ITO/SiO2 substrate from the GaAs wafer. Finally, the AlGaAs disks are detached from the PDMS stamp by applying heat to the thermal adhesive PPC layer.

**Optical experiments.** For nonlinear optical measurements, we pump the resonators with a tunable optical parametric amplifier (MIROPA-fs-M from Hotlight Systems) that generates 2 ps duration pulses at a repetition rate of 5.144 MHz pumped by a pulsed laser 1030 nm (FemtoLux by Ekspla). The considered spectral range of the pump is from 1500 to 1700 nm. To create vector beams over the pump spectral range, we use two custom q-plates optimized for 1570 nm and 1640 nm wavelengths (Thorlabs). To clear up the pump spectrum, we use a long-pass infrared filter (Thorlabs FELH1300). The polarization type of a pump beam is checked by full polarimetry performed with a combination of an achromatic quarter wave-plate and broadband wire-grid polarizer – Thorlabs AQWP05M-1600 and WP25M-UB, respectively. The pump beam is observed by a near-infrared InGaAs camera Xenics Bobcat-320 with a 150-mm focal distance achromatic doublet (AC508-150-CML). A train of wavelength tunable femtosecond laser pulses is focused from the air side of the sample mounted on a three-dimensional stage by a Mitutoyo MPlanApo NIR objective lens, ×100 infrared, 0.70 NA. The lens also collects the SHG in the backward direction. The incident pump and the backward generated second harmonic are separated by a 45° dichroic mirror. We measure the diameter of the azimuthally, radially, and linearly polarized focused pump laser beam by performing knife-edge experiments with details provided in Supplementary Information. The beam waist radius is 1.8 μm for the azimuthal pump. The fabricated nanoparticles are isolated from each other by large spacing of 10 μm. The SHG signal is collected by an Olympus objective lens MPlanFL N (×100 visible, 0.90 NA) and detected by a visible cooled CCD camera (Starlight Xpress Ltd, Trius-SX694) with a 150-mm focal distance achromatic doublet. The detected signal is filtered out by a set of filters (a coloured glass bandpass filter Thorlabs FGB25 and UV fused silica filter with dielectric coating FELH0650). The SHG signal is normalized over a spectral function of the setup which includes filter transmittance, laser power, and detector sensitivity spectra. The origin of the SHG signal is verified by the direct measurement of its spectrum (by a visible spectrometer Ocean Optics QE Pro) and its power dependence which is in a quadratic manner. The directionality diagrams of the second harmonic are measured by recording the back-focal plane images of the two objective lenses with an additional lens (75 mm focal distance achromatic doublet). Polarization states of the SHG over the directionality diagram are retrieved with Stokes vector formalism by employing an achromatic quarter-wave plate and a polarizer.

**Theoretical methods.** A detailed description of the derivation of Eq. (1) and the related method of the expansion of the Green's function over the eigenmodes are provided in Supplementary Information.

**Numerical calculations.** For numerical simulations, shown in Figs. 1 and 2, we use the finite-element-method eigenvalue solver in COMSOL Multiphysics. All calculations are realized for a single nanoantenna of a specific size on a semi-infinite structured substrate surrounded by a perfect matched layer mimicking an infinite region. All material properties including losses are imported from the tabulated data for AlGaAs (20% Al) and $SiO_2$ and extracted from the experimental ellipsometry data for the ITO layer. The measured Q factor is extracted from the experimental scattering spectrum using single-peak fitting to a Fano lineshape via the Levenberg-Marquardt algorithm. The error is due to the inaccuracy of the fitting procedure.

**Acknowledgements**

The authors thank D. Smirnova, L. Wang, and B. Luther-Davies for their valuable inputs into this project at various stages of its development. They also acknowledge a financial support from the Australian Research Council, the Strategic Fund of the Australian National University, and the Russian Science Foundation (grant 18-72-10140). K.K. and A.B. acknowledge a support from the Foundation for the Advancement of Theoretical Physics and Mathematics "BASIS". H.-G.P. acknowledges a support from the National Research Foundation of Korea (NRF) grant funded by the Korean Government (MSIP) (grant 2018R1A3A3000666).

**Author contributions**

K.K, S.K., and Y.K. conceived the research. K.K. and A.B. performed numerical simulations and theoretical calculations. J.-H.C. and H.-G.P fabricated the samples. S.K. and E.M.-G conducted experimental studies. K.K., S.K. and Y.K. wrote the manuscript based on the input from all the authors. All authors contributed to writing and editing the manuscript.

**Competing financial interests**

The authors declare no competing financial interests.

Correspondence and requests for materials should be addressed to Hong-Gyu Park or Yuri Kivshar.

# Supplementary Information.
# Individual nanoantennas empowered by bound states in the continuum for nonlinear photonics

Kirill Koshelev[1,2], Sergey Kruk[1], Elizaveta Melik-Gaykazyan[1,3], Jae-Hyuck Choi[4], Andrey Bogdanov[2], Hong-Gyu Park[4], Yuri Kivshar[1,2]

[1]*Nonlinear Physics Center, Australian National University, Canberra ACT 2601, Australia*

[2]*ITMO University, St. Petersburg 197101, Russia*

[3]*Faculty of Physics, Lomonosov Moscow State University, Moscow 119991, Russia*

[4]*Department of Physics, Korea University, Seoul 02841, Republic of Korea*

**Contents**

- Modes of an individual disk in free space and on a substrate;
- Linear simulations and multipolar decomposition;
- Nonlinear simulations and multipolar decomposition;
- Eigenmode expansion method for open optical resonators;
- Derivation of Equation 1 for the total second harmonic power;
- Analysis of the spatial coupling coefficient;
- Sample fabrication;
- Linear spectroscopy;
- Experimental setup for nonlinear spectroscopy;
- Knife-edge experiment;
- Peak power dependence of the measured SHG signal;
- Directivity diagram of the second harmonic;
- Comparison of SHG efficiencies from nanoscale dielectric and plasmonic resonators.

# 1. Modes of an individual disk in free space and on a substrate.

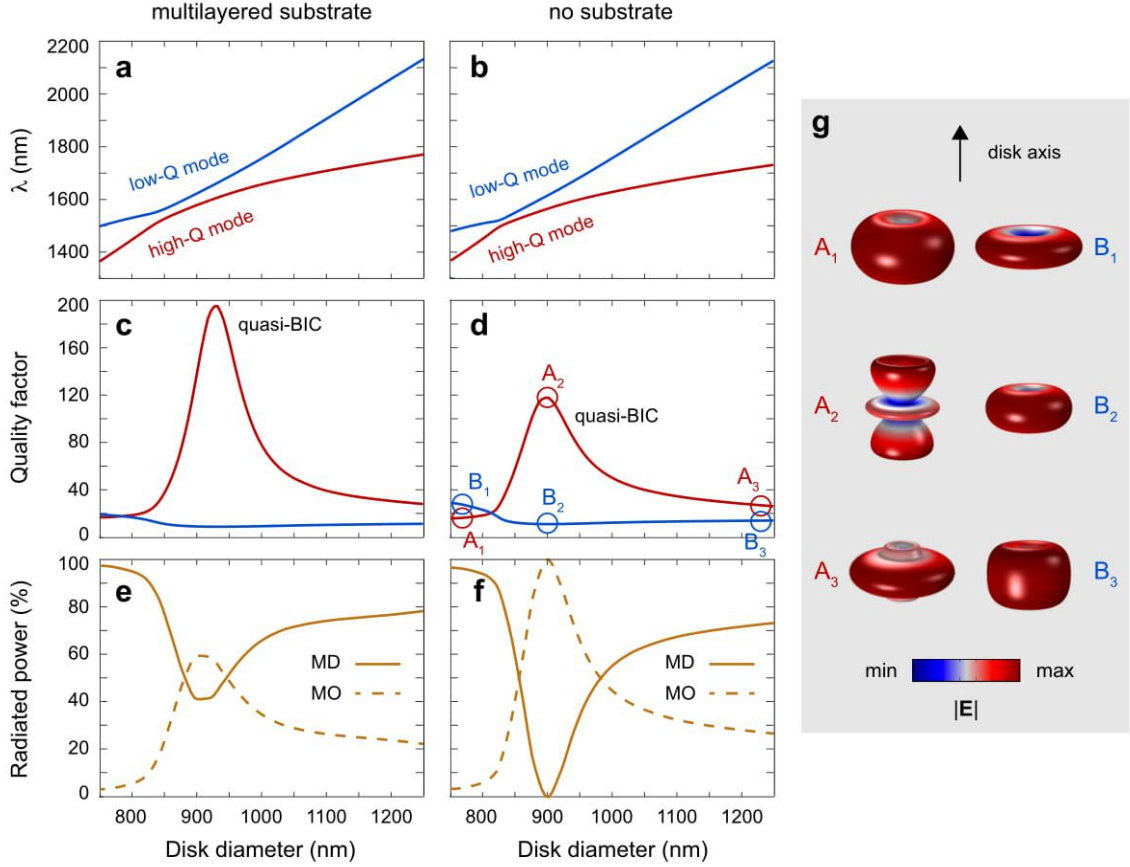

Figure 1: Behavior of eigenmodes in an individual disk resonator in the strong mode coupling regime (a, c, e) in free space and (b, d, f) on a three-layer (SiO2/ITO/glass) substrate with a 300 nm ITO layer and a 350 nm SiO$_2$ spacer. (a,b) Dispersion of the low-Q and high-Q modes vs. resonator diameter. (c,d) Q factor evolution for the low-Q and high-Q modes vs. resonator diameter. (e,f) Multipolar decomposition of the radiated power for the high-Q mode vs. resonator diameter. MD and MO denote the magnetic dipole and magnetic octupole contributions. (f) Transformation of the radiation pattern (far-field distribution) for the high-Q and low-Q mode while passing the avoided resonance crossing. Points $A_{1,2,3}$, $B_{1,2,3}$ correspond to the disk diameters of 750 nm, 900 nm, 1250 nm, respectively.

First, we explain the mechanism of formation of quasi-BICs in an individual subwavelength disk resonator. A disk supports two families of leaky modes, which can be approximately divided into radially and axially polarized modes. The eigenmodes of a disk can be classified according to their azimuthal order ($m$) with respect to the disk axis and their parity with respect to the up-down inversion. Any pair of radial and axial modes with the same azimuthal index and same parity are coupled if the difference between their frequencies is decreased. We select a pair of low-frequency radial and axial modes with ($m = 0$) and couple them via continuous tuning of the disk diameter.

For calculations, we consider two different cases: (i) the experimental design - a disk on top of a three-layer substrate (SiO2/ITO/glass) with a 300 nm ITO layer and a 350 nm SiO$_2$ spacer and (ii) a disk suspended in air without a substrate. As shown in Figs. 1a-d, coupling between the modes produces the characteristic avoided resonance crossing of frequency curves and modification of mode radiative Q factors. Figure 1g shows the transformation of the far-field patterns while passing the avoided resonance crossing. Away from the avoided resonance crossing each mode represents a magnetic dipole aligned with the disk axis. At the quasi-BIC regime the dipolar contribution to the radiation of the high-Q mode is suppressed because of destructive interference between two magnetic dipoles, therefore, the radiation pattern changes to a magnetic octupole.

For quantitative characterization of the disk eigenmodes, we employ the mode decomposition method over the irreducible spherical multipoles [1], characterized by both azimuthal (*m*) and orbital (*l*) indices. The decomposition is realized as a custom built-in routine for the eigenmode solver in COMSOL Multiphysics. For multipole classication, we use the following notations: electric/magnetic dipole (ED/MD, $l = 1$), electric/magnetic quadrupole (EQ/MQ, $l = 2$), electric/magnetic octupole (EO/MO, $l = 3$). We note that the multipolar decomposition for eigenmodes of open resonators is approximate due to divergence of field amplitudes in the far-field zone. Despite that, it still gives good quantitative results for modes with a Q factor more than 10.

Figures 1e and f show the multipolar decomposition of the radiated power for the high-Q mode. It can be seen, that in the vicinity of the quasi-BIC the magnetic dipole contribution is strongly suppressed and the magnetic octupole contribution dominates which is in agreement with the evolution of the far-field patterns shown in Fig. 1g. We note that the presence of a substrate breaks the spherical symmetry of the far-field zone and induces interference between different multipoles, which leads to some error in the results of the standard multipolar decomposition (for results shown in Fig. 1e).

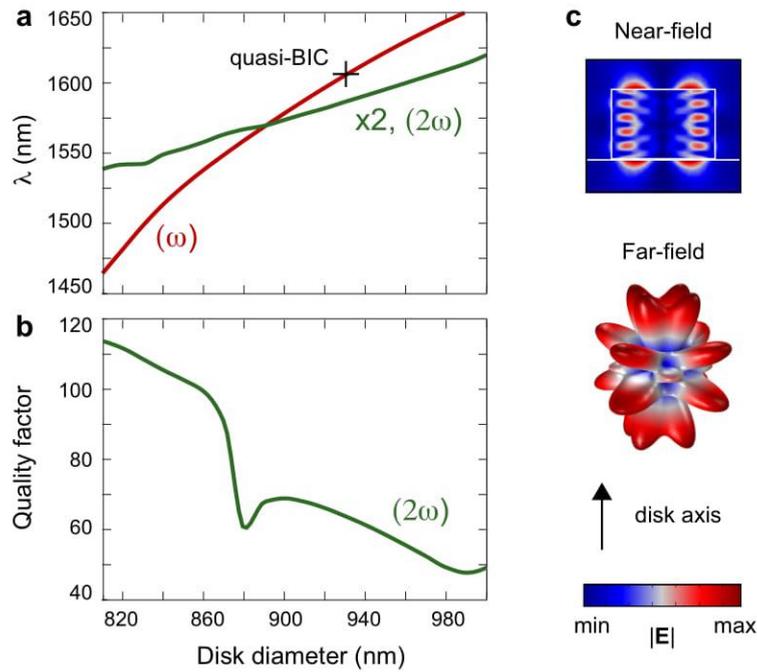

Figure 2: Mode dispersion at the second harmonic wavelength. (a) Dispersion of the high-Q mode at the pump wavelength (red) and the mode at the second harmonic wavelength (green) vs. resonator diameter. The wavelength of the SH mode is doubled for a direct comparison. (b) Q factor evolution for the mode at the second harmonic wavelength vs. resonator diameter. (c) Near-field and far-field patterns for mode at the second harmonic wavelength, calculated for a disk without a substrate.

At the second harmonic wavelength, in the vicinity of the doubled frequency of the quasi-BIC the disk supports a single high-order eigenmode. Figure 2 shows the mode dispersion [panel (a)], the Q factor evolution [panel (b)] and the field profiles for the second harmonic mode [panel (c)]. At the quasi-BIC regime, the second harmonic mode has a Q factor of about 65.

## 2. Linear simulations and multipolar decomposition.

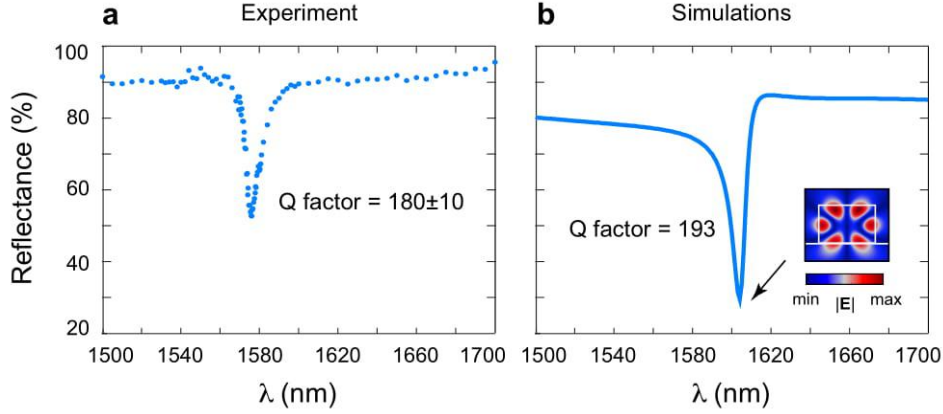

Figure 3: Reflectance spectrum and retrieved Q factor of the observed resonance for a 930 nm disk on a three-layer substrate excited with an azimuthal pump. (a) Measurements. (b) Simulations. The inset shows the near-field pattern at the resonant wavelength.

For numerical simulations of the linear spectrum we use the finite-element-method solver in COMSOL Multiphysics in the frequency domain. To compare the simulations with the measured spectra, we consider an AlGaAs disk on top of a three-layer substrate (SiO2/ITO/glass) with a 350 nm $SiO_2$ spacer and the azimuthally polarized pump. To obtain the exact expression for the background field for scattering simulations, we derive the angular spectrum representation for the azimuthal cylindrical vector beam in each layer of the multilayered structure and match the solutions at the boundaries between the layers.

We calculate the reflected power coming through 0.7NA aperture in the backward direction and normalise it on the reflectance of the multilayered substrate. The comparison between the measured and simulated reflectance is shown in Fig. 3. We observe that the dip in the simulated spectrum is red-shifted by 28 nm with respect to the dip observed in the experiment. This difference can be explained by unknown experimental data for the permittivity of AlGaAs and fabrication imperfections such as inclination of the disk walls, which are not taken into account in the simulations.

To understand the nature of the resonant dip in the reflectance spectrum, we perform the multipolar decomposition of the total scattered power, shown in Fig. 4. It demonstrates, that the quasi-BIC is determined by the magnetic dipolar and octupolar patterns in accordance to the results of the eigenmode simulations (see Fig. 1e).

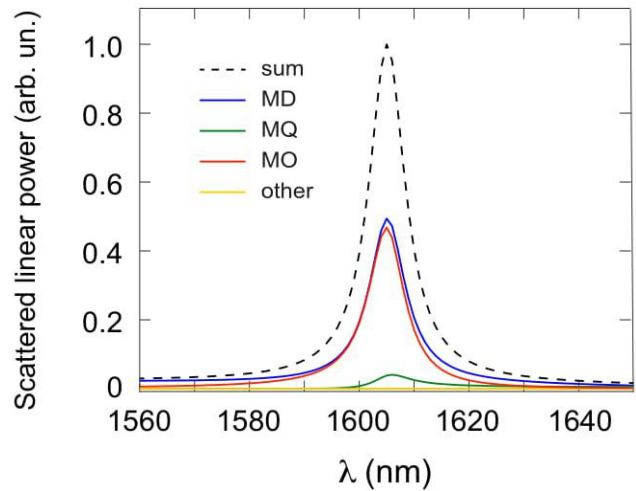

Figure 4: Multipolar decomposition of the scattered power at the pump wavelength.

## 3. Nonlinear simulations and multipolar decomposition.

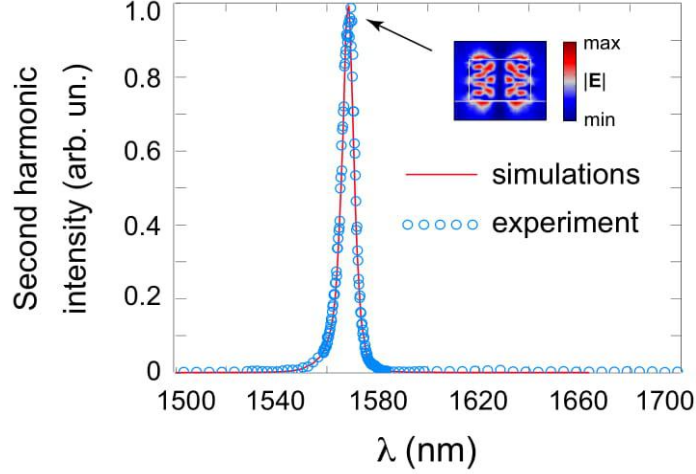

Figure 5: Experimental and simulated spectrum of the second harmonic intensity for a 930 nm disk on a three-layer substrate excited with an azimuthal pump. The simulated spectrum is artificially blue-shifted by 34 nm. The simulated and experimental data are normalized independently. The inset shows the near-field pattern at the resonant peak wavelength.

For numerical simulations of the nonlinear response at the second-harmonic wavelength, we employ the approach based on the undepleted pump approximation, using two steps to calculate the intensity of the radiated nonlinear signal. Using the simulated field amplitudes at the pump wavelength, we obtain the nonlinear polarization induced inside the disk. Then, we employ the polarization as a source for the electromagnetic simulation at the harmonic wavelength to obtain the generated second-harmonic field. The nonlinear susceptibility tensor corresponds to the zincblende crystalline structure with $\chi^{(2)}_{xyz} = 290$ pm/V.

We calculate the sum of the radiated second harmonic intensity coming through 0.7NA aperture in the backward direction and 0.9NA aperture in the forward direction. A comparison between the measured and simulated second harmonic intensity is shown in Fig. 5. The difference in the peak positions can be explained by unknown experimental data for the permittivity of AlGaAs and fabrication imperfections such as inclination of the disk walls, which are not taken into account in the simulations. Figure 6 demonstrates the multipolar decomposition of the total second harmonic power which shows that the resonant peak is strongly dominated by the magnetic quadrupolar mode.

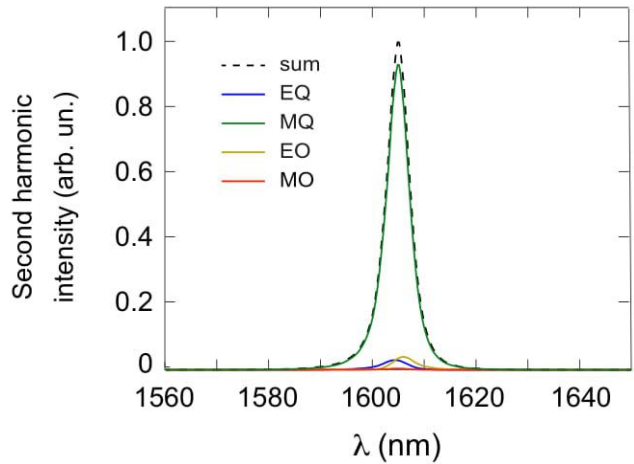

Figure 6: Multipolar decomposition of the total second harmonic intensity.

## 4. Eigenmode expansion method for open optical resonators.

We present a brief introduction to the eigenmode expansion method for non-Hermitian optical systems. The theory was first developed for description of optical properties of one-dimensional photonic crystal slabs [2], where the eigenmode expansion of the Green's function was derived in the general form. Later, the theory was extended for optical resonators of arbitrary shape and dimension [3, 4], including magnetic, chiral and bianisotropic materials with the frequency dispersion [5].

We study an open optical resonator, which possesses properties of a non-Hermitian system and its modes are inherently leaky. The main idea is to expand the resonator's Green's function $\hat{\mathbf{G}}$ over the eigenmodes $\mathbf{E}_j(\mathbf{r})$

$$\hat{\mathbf{G}}(\omega, \mathbf{r}, \mathbf{r}') = \sum_j c^2 \frac{\mathbf{E}_j(\mathbf{r}) \otimes \mathbf{E}_j(\mathbf{r}')}{2N_j \omega(\omega - \omega_j + i\gamma_j)}. \tag{1}$$

Here, $\omega_j$ and $\gamma_j$ are the real and imaginary parts of the eigenfrequencies, respectively, and $N_j$ is the normalisation constant for $\mathbf{E}_j$. The expansion holds only for $\mathbf{r}, \mathbf{r}'$ inside the resonator, where the permittivity and permeability are different from the background values. The eigenfunctions are diverging in space for large arguments $|\mathbf{r}| \gg \lambda$, which is the result of their leaky nature. The main difficulty of such approach is the accurate and correct normalisation of eigenmodes, which was proposed only recently [4]

$$N_j = \int_V dV \, \varepsilon \mathbf{E}_j \cdot \mathbf{E}_j + \frac{c^2}{2(\omega_j - i\gamma_j)^2} \oint_S dS \left[ \mathbf{E}_j \cdot \frac{\partial}{\partial r} r \frac{\partial \mathbf{E}_j}{\partial r} - r \left( \frac{\partial \mathbf{E}_j}{\partial r} \right)^2 \right], \tag{2}$$

where both volume and surface integration goes over the spherical shell $S$ located in the far field zone.

Using the expression Eq. 1 for the Green's function it is possible to find the solution of Maxwell's equations for any given source. The eigenfunctions can be found numerically using full-wave computational packages, or semi-analytically, using the resonant state expansion method [2].

## 5. Derivation of Equation 1 for the total second harmonic power.

*Linear response*

We consider a subwavelength AlGaAs disk on a glass substrate. We neglect non-radiative losses, since the non-radiative Q factor $\text{Re}[\varepsilon]/\text{Im}[\varepsilon] > 10^3$ within the range of wavelengths from 780 to 1670 nm. The disk is excited by the pump with a given distribution of electric field $\mathbf{E}_{\text{bg}}$. The total electric field $\mathbf{E}(\mathbf{r})$ can be divided into the background and the scattered as

$$\mathbf{E}(\omega, \mathbf{r}) = \mathbf{E}_{\text{sc}}(\omega, \mathbf{r}) + \mathbf{E}_{\text{bg}}(\omega, \mathbf{r}). \tag{3}$$

The scattered field can be found using the Green's function

$$\mathbf{E}_{\text{sc}}(\omega, \mathbf{r}) = -\frac{\omega^2}{c^2} \int d\mathbf{r}' \, \Delta\epsilon(\omega, \mathbf{r}') \hat{\mathbf{G}}(\omega, \mathbf{r}, \mathbf{r}') \cdot \mathbf{E}_{\text{bg}}(\omega, \mathbf{r}'), \tag{4}$$

where $\Delta\epsilon(\omega, \mathbf{r}) = \epsilon(\omega, \mathbf{r}) - \epsilon_{\text{bg}}(\omega, \mathbf{r})$.

The scattered field can be rigorously expanded over the eigenmodes of the resonator. We consider a case when only one mode $\mathbf{E}_1$ is resonantly excited in the vicinity of the pump frequency

$$\mathbf{E}_{\text{sc}}(\omega, \mathbf{r}) = a(\omega) \mathbf{E}_1(\mathbf{r}), \tag{5}$$

Using the expansion in Eq. 1 we find the resonant amplitude $a$ at the pump wavelength

$$a(\omega) = -\frac{\omega}{2N_1(\omega - \omega_1 + i\gamma_1)} \int d\mathbf{r}' \, \Delta\epsilon(\omega, \mathbf{r}') \mathbf{E}_1(\mathbf{r}') \cdot \mathbf{E}_{\text{bg}}(\omega, \mathbf{r}'). \tag{6}$$

The energy accumulated inside the resonator $W(\omega)$ is proportional to $|a|^2$, which can be re-written as

$$W(\omega) \propto |a(\omega)|^2 = \frac{c}{N_1 \omega_1} Q_1 L_1(\omega) \varkappa_1(\omega) P_0'(\omega). \tag{7}$$

Here, $Q_j = \omega_j/2\gamma_j$ is the mode quality factor, the spectral overlap factor $L_j(\omega)$ is

$$L_j(\omega) = \frac{\gamma_j^2}{(\omega - \omega_j)^2 + \gamma_j^2}, \tag{8}$$

and the coupling coefficient $\varkappa_1$ is

$$\varkappa_1(\omega) = \frac{\left|(\omega/c) \int d\mathbf{r}' \, \Delta\epsilon(\omega, \mathbf{r}') \mathbf{E}_1(\mathbf{r}') \cdot \mathbf{E}_{\text{bg}}(\omega, \mathbf{r}')\right|^2}{(2\gamma_1/c) N_1 P_0'(\omega)}. \tag{9}$$

The coefficient $P_0'$ is proportional to the total incident power $P_0$

$$P_0'(\omega) = \frac{8\pi}{c} P_0. \tag{10}$$

*Nonlinear response*

To analyse the resonator response at the second harmonic (SH) wavelength we calculate the nonlinear SH polarization

$$P_i(2\omega) = \sum_{j,k} \chi^{(2)}_{ijk} E_j(\omega) E_k(\omega), \tag{11}$$

where $\chi^{(2)}_{ijk}$ is the second-order susceptibility tensor. For AlGaAs $\chi^{(2)}$ has the symmetry of zincblende crystalline structure. We assume nearly resonant conditions when the amplitude $a(\omega)$ is large, so $\mathbf{E}(\omega, \mathbf{r}) \simeq \mathbf{E}_{\text{sc}}(\omega, \mathbf{r})$ and

$$P_i(2\omega) = [a(\omega)]^2 \sum_{j,k} \chi^{(2)}_{ijk} E_{1,j} E_{1,k}. \tag{12}$$

The induced field $\mathbf{E}(2\omega)$ at the SH wavelength can be found using the resonator Green's function similar to Eq. 4

$$\mathbf{E}(2\omega, \mathbf{r}) = -\frac{(2\omega)^2}{c^2} \int d\mathbf{r}' \hat{\mathbf{G}}(2\omega, \mathbf{r}, \mathbf{r}') \cdot \mathbf{P}(2\omega, \mathbf{r}'). \tag{13}$$

We assume that $\mathbf{E}(2\omega)$ is dominated by a single resonant state $\mathbf{E}_2$ with frequency $\omega_2$ lying in the vicinity of $2\omega$

$$\mathbf{E}(2\omega, \mathbf{r}) = b(2\omega)\mathbf{E}_2(\mathbf{r}). \tag{14}$$

Thus, the amplitude $b$ can be found as

$$b(2\omega) = -\frac{2\omega}{2N_2(2\omega - \omega_2 + i\gamma_2)} \int d\mathbf{r}\, \mathbf{E}_2(\mathbf{r}) \cdot \mathbf{P}(2\omega, \mathbf{r}) = -\frac{2\omega\,[a(\omega)]^2}{2N_2(2\omega - \omega_2 + i\gamma_2)} \sum_{i,j,k} \chi^{(2)}_{ijk} \int d\mathbf{r}\, E_{2,i}E_{1,j}E_{1,k}. \tag{15}$$

The energy of the SH field is proportional to $|b(2\omega)|^2$

$$|b(2\omega)|^2 = \frac{(2\omega/c)^2}{(2\gamma_2/c)N_2} Q_2 L_2(2\omega) \varkappa_{12} \left[\frac{\omega_1}{c} N_1 |a(\omega)|^2\right]^2, \tag{16}$$

where the cross-coupling coefficient $\varkappa_{12}$ is

$$\varkappa_{12} = \frac{\left|\sum_{i,j,k} \chi^{(2)}_{ijk} \int d\mathbf{r}\, E_{2,i}(\mathbf{r})E_{1,j}(\mathbf{r})E_{1,k}(\mathbf{r})\right|^2}{(N_2\omega_2/c)(N_1\omega_1/c)^2}. \tag{17}$$

The total SH power

$$P(2\omega) = \frac{c}{8\pi} \oint_S d\mathbf{S} \cdot \mathrm{Re}\left[\mathbf{E}(2\omega) \times \mathbf{H}^*(2\omega)\right]. \tag{18}$$

Here, the integral is evaluated at the disk surface, where Eq. 14 is valid.

Finally, combining Equations (7), (14), (16) and (18), we get the expression for the total SH power

$$P(2\omega) = \alpha(2\omega)\varkappa_2 Q_2 L_2(2\omega) \varkappa_{12} \left[Q_1 L_1(\omega) \varkappa_1(\omega) P_0(\omega)\right]^2. \tag{19}$$

Here, the decoupling coefficient $\varkappa_2$ is

$$\varkappa_2 = \frac{\oint_S d\mathbf{S} \cdot \mathrm{Re}\left[\mathbf{E}_2 \times \mathbf{H}_2^*\right]}{(2\gamma_2/c)N_2}, \tag{20}$$

and the smooth envelope coefficient $\alpha$ is

$$\alpha(2\omega) = \frac{8\pi}{c}\left[\frac{2\omega}{c}\right]^2. \tag{21}$$

## 6. Analysis of the spatial coupling coefficient.

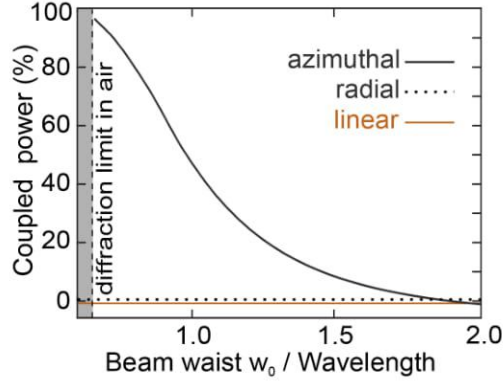

Figure 7: Percent of the pump power coupled to the quasi-BIC depending on the ratio between the beam waist and the pump wavelength. The calculation is done for a disk surrounded by air. Different curves show azimuthally, radially, and linearly polarized cylindrical pump beams.

In the previous section we derived the expression Eq. 9 for the spatial coupling coefficient $\varkappa_1$ between the pump and the resonator mode at the pump wavelength. The volume integral from the numerator can be recast to the surface integral in the vicinity of resonance $\omega \simeq \omega_1$ ($\gamma_1 \ll \omega_1$) using the divergence theorem [4]

$$\frac{\omega_1}{c} \int d\mathbf{r}\, \Delta\epsilon(\omega_1,\mathbf{r}) \mathbf{E}_1(\mathbf{r}) \cdot \mathbf{E}_{\text{bg}}(\omega_1,\mathbf{r}) = \frac{c}{\omega_1} \oint_{S_0} dS \left[ \mathbf{E}_1(\mathbf{r}) \cdot \frac{\partial}{\partial r_n} \mathbf{E}_{\text{bg}}(\omega_1,\mathbf{r}) - \mathbf{E}_{\text{bg}}(\omega_1,\mathbf{r}) \cdot \frac{\partial}{\partial r_n} \mathbf{E}_1(\mathbf{r}) \right], \qquad (22)$$

where $r_n$ is the component of $\mathbf{r}$ normal to the resonator's surface $S_0$. Using the same divergence theorem, the mode inverse radiation lifetime can be expressed as

$$\frac{2\gamma_1}{c} \int d\mathbf{r}\, \epsilon(\omega_1,\mathbf{r}) |\mathbf{E}_1(\mathbf{r})|^2 = \frac{c}{2\omega_1} \oint_S dS \left[ \mathbf{E}_1(\mathbf{r}) \cdot \frac{\partial}{\partial n} \mathbf{E}_1^*(\mathbf{r}) - \mathbf{E}_1^*(\mathbf{r}) \cdot \frac{\partial}{\partial n} \mathbf{E}_1(\mathbf{r}) \right] = \oint_S dS\, |\mathbf{E}_1(\mathbf{r})|^2, \qquad (23)$$

for the surface $S$ located in the far field zone as in Eq. 2.

Finally, the expression for $\varkappa_1$ at the resonant conditions can be written as

$$\varkappa_1(\omega_1) = \xi \frac{\left| c/\omega_1 \oint_{S_0} dS \left[ \mathbf{E}_1(\mathbf{r}) \cdot \frac{\partial}{\partial r_n} \mathbf{E}_{\text{bg}}(\omega_1,\mathbf{r}) - \mathbf{E}_{\text{bg}}(\omega_1,\mathbf{r}) \cdot \frac{\partial}{\partial r_n} \mathbf{E}_1(\mathbf{r}) \right] \right|^2}{\oint_S dS\, |\mathbf{E}_1(\mathbf{r})|^2 \int_{S_\infty} d\mathbf{S} \cdot \text{Re}\left[ \mathbf{E}_{\text{bg}}(\omega_1,\mathbf{r}) \times \mathbf{H}_{\text{bg}}^*(\omega_1,\mathbf{r}) \right]}. \qquad (24)$$

Here, the coefficient $\xi$ shows how leaky the mode is

$$\xi = \frac{\int d\mathbf{r}\, \epsilon(\omega_1,\mathbf{r}) |\mathbf{E}_1(\mathbf{r})|^2}{N_1}, \qquad (25)$$

which is $\xi \simeq 1$ for high-Q modes.

We calculate $\varkappa_1$ for the quasi-BIC for azimuthally, radially and linearly polarized cylindrical pump beam depending on the beam waist $w_0$ (see Fig. 7 and Fig.2 c of the main text). For radially and linearly polarized beams the coupling coefficient is exactly zero. For azimuthally polarized beam $\varkappa_1$ is very high even for weakly focused beams $w_0 > \lambda$, despite the pump electric field goes to zero at the disk center. This is explained by the numerator of Eq. 24 showing that $\varkappa_1$ depends on the pump field distribution on the entire surface of the resonator which is determined also by the longitudinal magnetic field component.

Equation 24 can be simplified assuming (i) the background field is close to a linearly-polarized plane wave, (ii) the surface $S_0$ is the surface of the cylindrical resonator with top and bottom surfaces perpendicular to the propagation of the incident wave, (iii) the mode electric field $\mathbf{E}_1$ has one dominant component aligned with the the pump field polarization, (iv) the mode is radiated mostly along the axis of the propagation of the incident wave, (v) the fields $\mathbf{E}_1$ and $\mathbf{E}_{bg}$ are almost uniform on the top and bottom surface of $S_0$ and (vi) the surface $S_0$ is effectively in the far-field of the mode. Then,

$$\varkappa_1(\omega_1) = \frac{2P_0^{\text{top}}}{P_0}, \qquad (26)$$

where $P_0^{\text{top}}$ is the power of the pump coming through the top surface of the resonator.

Despite the definition, Eq. 26 requires additional assumptions, it was widely used for evaluation of efficiency of nonlinear processes, especially for Gaussian vector beam pumps (see Refs. 26-27 of the main text). We will use this definition for a comparison with earlier results in Section 13.

## 7. Sample fabrication.

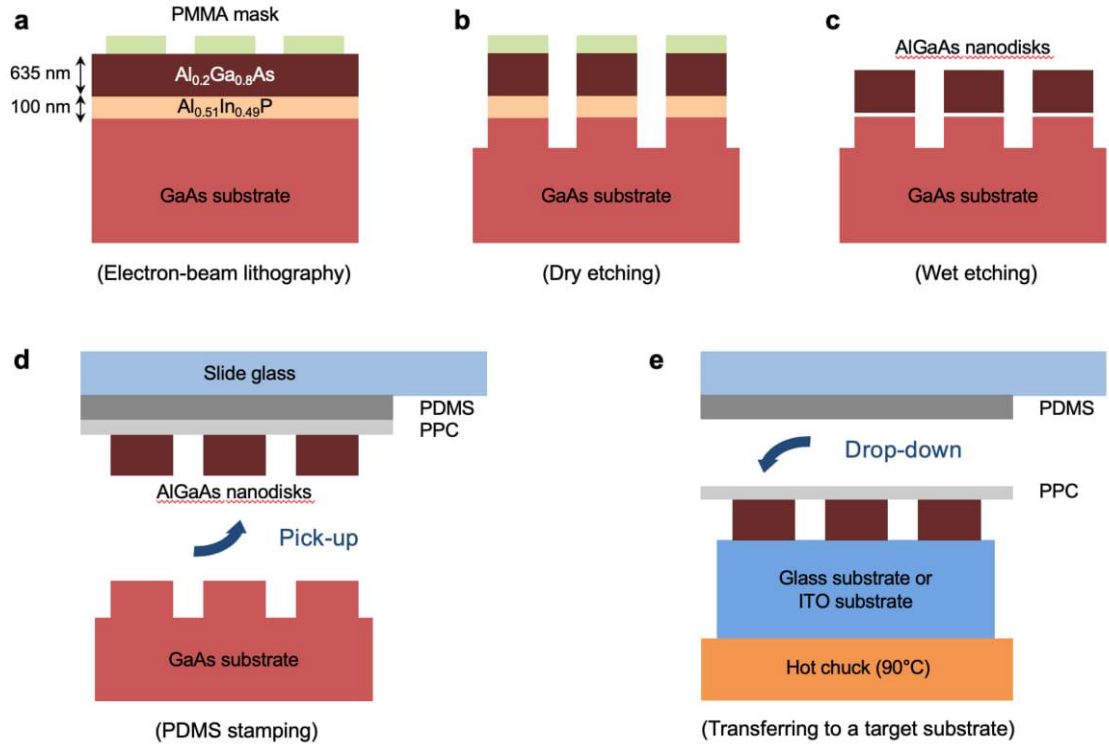

Figure 8: Schematics of the fabrication procedure. (a) PMMA mask is defined on top of the AlGaAs/AlInP/GaAs wafer structure using electron-beam lithography. (b) Vertical pillar structure is fabricated by chemically-assisted ion beam etching. (c) The PMMA mask is removed by $O_2$ plasma. The AlInP buffer layer is selectively wet-etched by diluted HCl solution. Then, the AlGaAs nanodisks are put on the GaAs substrate. (d) The AlGaAs nanodisks are picked up from the GaAs substrate using a polypropylene carbonate (PPC)-coated polydimenthylsiloxane (PDMS) stamp. (e) The nanodisks are dropped down to a target substrate such as glass or ITO substrates by applying heat (up to 90°C) to the thermal adhesive PPC layer. The PPC layer is then removed thoroughly by acetone.

## 8. Linear spectroscopy.

The schematic of the experimental setup for linear spectroscopy by using pulsed laser radiation generated by a tunable optical parametric amplifier (MIROPA-fs-M from Hotlight Systems) is presented in Fig. 9.

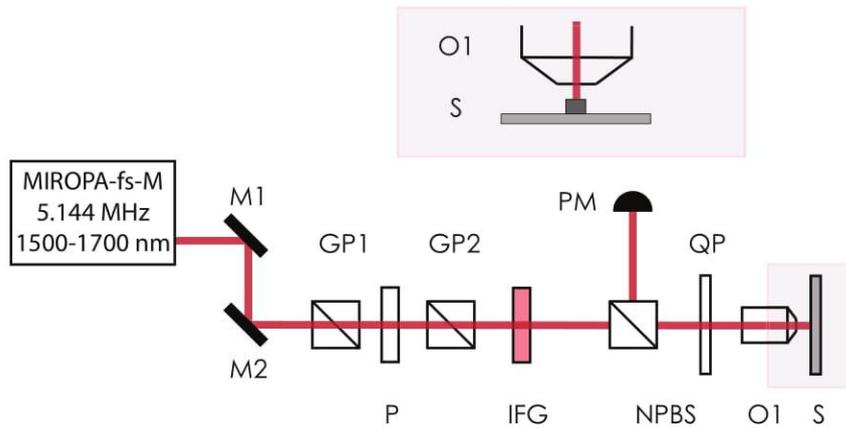

Figure 9: Experimental setup for linear spectroscopy in reflection. MIROPA-fs-M is the optical parametric amplifier, M1 and M2 are the mirrors, GP1 and GP2 are the Glan prisms, P is the wire-grid polarizer, IFG is the infrared glass filter, QP is the commercial liquid crystal q-plate, NPBS is the non-polarizing 50/50 beamsplitter, O1 is the objective, S is the sample on a three-dimensional stage, PM is the power meter head.

Raw data of the reflected power spectra are shown in Fig. 10.

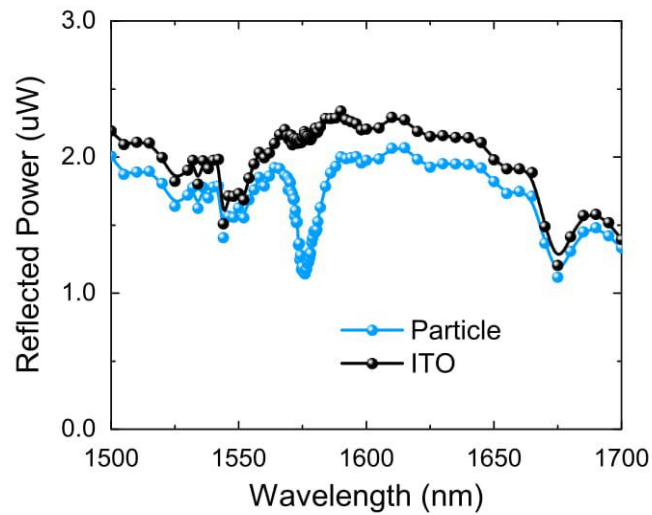

Figure 10: Spectra of the power reflected by an AlGaAs nanoresonator (blue dots) and by an ITO substrate (black dots). Radiation is collected in the backward direction by an objective lens with 0.70 NA.

## 9. Experimental setup for nonlinear spectroscopy.

The schematic of the experimental setup for second-harmonic generation (SHG) spectroscopy is presented in Fig. 11.

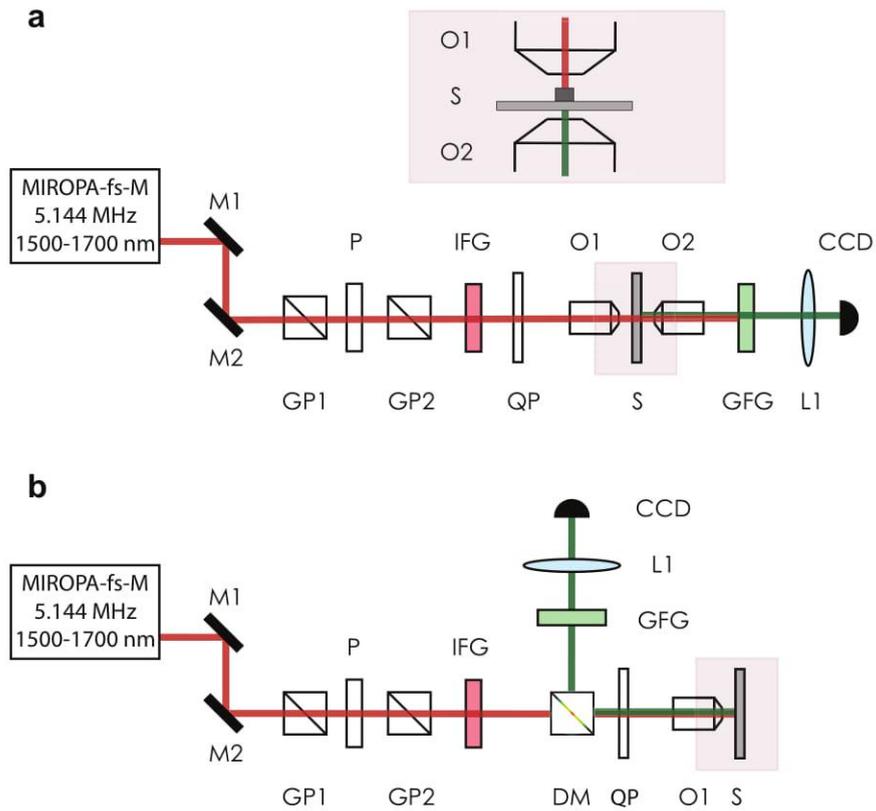

Figure 11: SHG experimental setup in (a) transmission and (b) reflection. MIROPA-fs-M is the optical parametric amplifier, M1 and M2 are the mirrors, GP1 and GP2 are the Glan prisms, P is the wire-grid polarizer, IFG is the infrared glass filter, QP is the commercial liquid crystal q-plate, O1 and O2 are the objectives, S is the sample on a three-dimensional stage, GFG is the set of filters which transmit the SHG signal, L1 is the lens, CCD is the detecting visible camera, DM is the dichroic mirror.

The detector counts are recalculated to the second harmonic optical signal power values using the quantum efficiency data for the Trius-SX694 CCD camera (Starlight Xpress Ltd). The calibration coefficient is also estimated by the use of a power meter and CW-laser.

## 10. Knife-edge experiment.

The geometrical parameters of cylindrical vector pump beams are measured using the knife-edge method [6]. The experimental results at the pump wavelength of 1640 nm for both linearly and azimuthally polarized pump beams are shown in Fig. 12. The considered parameter of the linearly polarized beam is the radius $w$ indicated by the level of $1/e^2$ of the peak value from the data derivative approximated by a Gaussian profile. The parameter $w_0$ defining an azimuthally polarized beam is estimated as the distance between two peaks of a derivative of beam profiling data because the derivative as a function of knife-edge position $x$ is proportional to the expression $(1 + 4x^2/w_0^2)e^{-2x^2/w_0^2}$.

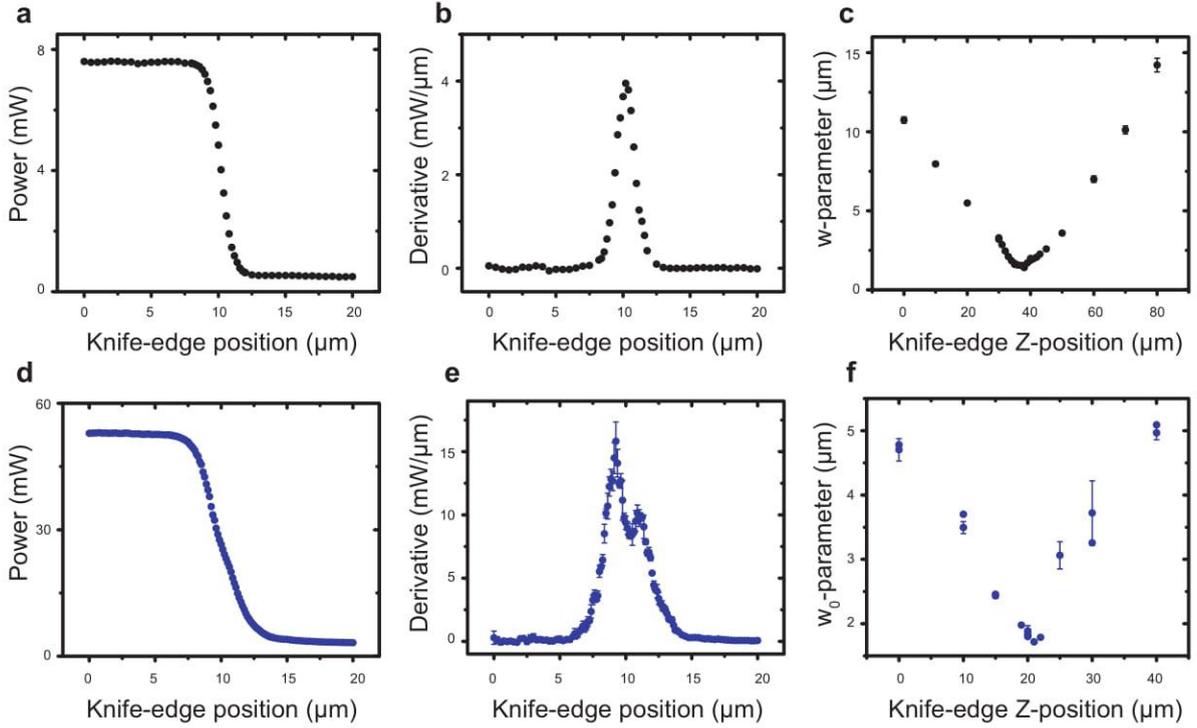

Figure 12: Knife-edge experimental results for linearly (a, b, c) and azimuthally (d, e, f) polarized focused beams. Power dependence on the knife-edge position (a, d) and corresponding derivatives of beam profiling data (b, e). Beam parameters $w$ (c) and $w_0$ (f) in dependence on the knife-edge position on the optical axis.

## 11. Peak power dependence of the measured SHG signal

Using an attenuator (a set of polarizing optical elements) we are able to conduct measurements of the second harmonic response with respect to the pump power. The experimental results are presented in Fig. 13 for the dependence of the second harmonic peak power $P_{p,\text{SH}}$ on the pump peak power $P_{p,0}$ (see Table 1 for details). As a considered particle is pumped, the resonant nonlinear response is blue-shifted.

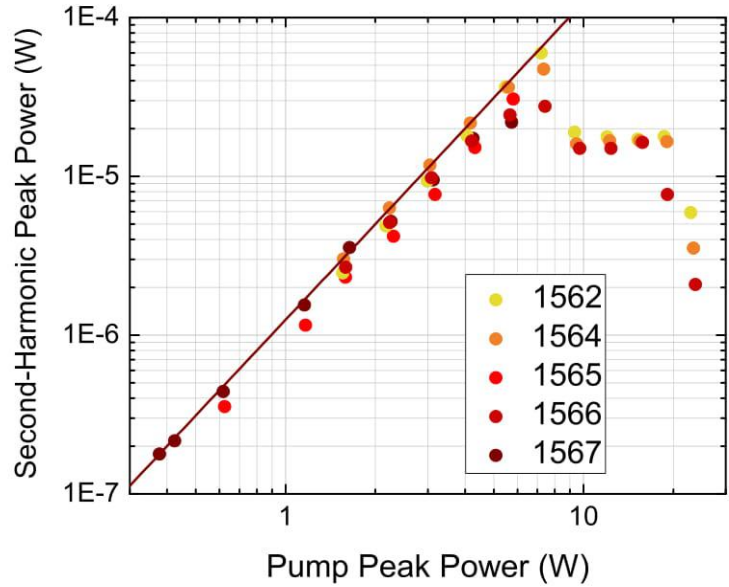

Figure 13: Peak power dependence of the total SHG signal generated by an AlGaAs nanoresonator (its height is 635 nm, diameter – 930 nm). Dot colors correspond to different central fundamental wavelengths of 1562, 1564, 1565, 1566, 1567 nm. Polarization of a pump beam is azimuthal. The brown line depicts a quadratic fit $P_{p,\text{SH}} = 1.3 \cdot 10^{-6} P_{p,0}^2$

## 12. Directivity diagram of the second harmonic.

By adding an additional lens we obtain back-focal plane images of the SHG signal. An achromatic quarter-wave plate and a wire-grid polarizer are implemented into the experimental setup to retrieve polarization states of the nonlinear response on the base of Stokes vector formalism. The obtained directivity diagrams are presented in Fig. 14.

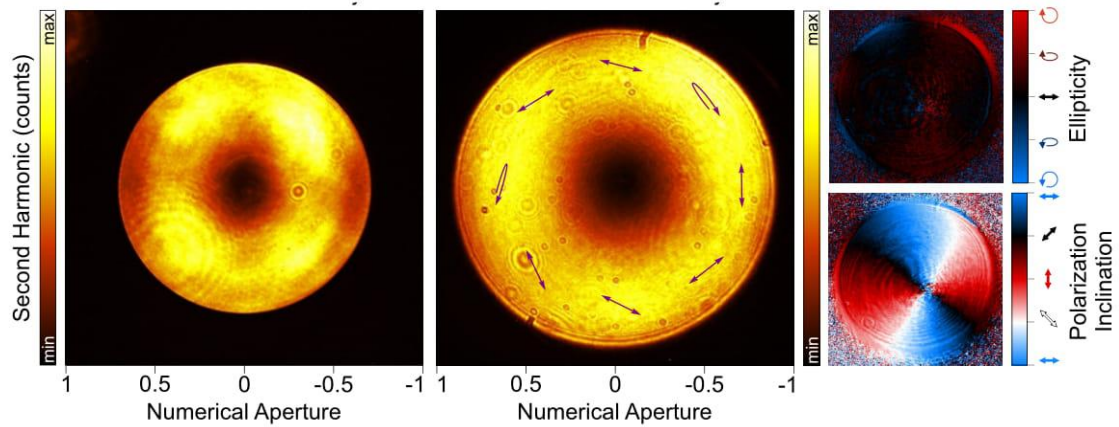

Figure 14: Experimentally measured directionality diagrams of the SHG in the backward (left) and forward (right) directions. Arrows visualize the polarization states. The insets show retrieved spatially-resolved ellipticity and polarization inclination angles for the SHG signal.

## 13. Comparison of SHG efficiencies from nanoscale dielectric and plasmonic resonators.

| Parameter | Pump $\lambda$ (nm) | Laser repetition rate (MHz) | Laser pulse duration (ps) | Excitation objective NA | Pump beam waist ($\mu$m) | Resonator radius (nm) | Pump average power (mW) | Pump peak power (W) | SH average power (nW) | SH peak power (mW) | Coupling coefficient (%) | Collection efficiency (%) | Measured conversion efficiency (W$^{-1}$) | Estimated total conversion efficiency (W$^{-1}$) |
|---|---|---|---|---|---|---|---|---|---|---|---|---|---|---|
| Symbol | $\lambda$ | $\nu$ | $\tau$ | NA | $w_0$ | $r_0$ | $P_0$ | $P_{p,0}$ | $P_{SH}$ | $P_{SH,0}$ | $\varkappa_1$ | $\beta$ | $\eta_{exp}$ | $\eta_{est}$ |
| Formula | | | | | $0.61\lambda/\mathrm{NA}$ | | | $0.94 P_0/\nu\tau$ | | $\frac{0.94 P_{SH}}{\sqrt{2}\nu\tau}$ | Eq. (29) | | Eq. (27) | Eq. (28) |
| **This work** | 1570 | 5.144 | 2 | 0.7 | 1.8* | 465 | $7.9\cdot10^{-2}$ | 7.2 | 0.93 | $6\cdot10^{-2}$ | 33** | 24 | $(1.3\cdot10^{-6})$*** | $4.8\cdot10^{-5}$ |
| **Ref. [27]** | 1556 | 5 | 0.5 | 0.85 | 1.1 | 245 | 1.0 | 376 | 8 | 2.1 | 19 | 30 | $1.5\cdot10^{-8}$ | $1.4\cdot10^{-6}$ |
| **Refs. [28,29]** | 1556 | 80 | 0.17 | 0.85 | 1.1 | 193 | 0.75 | 52 | 0.7 | $3.4\cdot10^{-2}$ | 12 | 10 | $1.3\cdot10^{-8}$ | $8.7\cdot10^{-6}$ |
| **Ref. [30]** | 910 | 0.1 | 0.18 | 0.9 | 0.62 | 200 | $5\cdot10^{-3}$ | 261 | $1\cdot10^{-2}$ | 0.42 | 38 | 28 | $6.2\cdot10^{-9}$ | $1.5\cdot10^{-7}$ |
| **Ref. [19]** | 1550 | 80 | 0.12 | 1.35 | 0.7 | N/A | 0.12 | 11.8 | $7.6\cdot10^{-4}$ | $5.3\cdot10^{-5}$ | N/A | N/A | $3.8\cdot10^{-10}$ | N/A |

Table 1: Experimental parameters for the comparison of SHG efficiencies from nanoscale dielectric and plasmonic resonators.
*Knife-edge measurements (see Section 10).
** Evaluated using Eq. 9.
*** Evaluated from the quadratic fit to the data shown in Fig. 4 of the main text

To compare the observed SHG conversion efficiency for our structure with previously demonstrated results in Refs. [19,27-30] of the main text, we evaluate and combine all the relevant parameters taking them directly from Refs. They are summarized in Table 1 which contains the data on the experimental setup properties. We note that in our experiments we measure the spot radius using the knife-edge method (see Section 10), while for other Refs. it was estimated using the diffraction limit criterion.

Table 1 summarizes the directly measured pump and second harmonic powers. The value of SH power is already renormalised to take into account the optical throughput of the optics and the detector losses, but do not account for collection efficiency of receiving objective. The peak powers are evaluated considering the Gaussian pulse shape, which gives the factor of 0.94. We assume the second harmonic pulse is $\sqrt{2}$ longer than the pump pulse, which is an approximation for a non-resonant bulk material.

For direct comparison of our results with the earlier works we use the directly measured value of the dimensional conversion efficiency, defined as

$$\eta_{\exp} = \frac{P_{p,\text{SH}}}{P_{p,0}^2}. \tag{27}$$

To evaluate and compare the intrinsic performance of the mode, independent on the pump coupling and the objective collection efficiency, we estimate the total SHG conversion efficiency. Its value is given by the ratio of the total radiated second-harmonic power estimated using the collection efficiency of the objective $\beta$ and the amount of the pump power coupled to the resonator proportional to the coupling coefficient $\varkappa_1$

$$\eta_{\text{est}} = \frac{\eta_{\exp}}{\beta \varkappa_1^2}. \tag{28}$$

To calculate the coupling coefficient for this work $\varkappa_1$ we use the exact expression [see Eq. (9)]. However, for other Refs. the numerical data for field profiles is not available, thus we use the simpified definition for $\varkappa_1$ [see Eq. (26))], evaluating it as the doubled percentage of pump power coming through the top surface of the resonator. For the linearly polarized fundamental Gaussian vector beam (for Refs. [19,27-30]) focused on the top surface of the disk resonator with radius $r_0$ the coupling efficiency can be calculated analytically

$$\varkappa_1 = 2\left[1 - \exp\left(-2r_0^2/w_0^2\right)\right]. \tag{29}$$

The collection efficiency $\beta$ is calculated as a percentage of the second harmonic power coming through the collecting objective, which is estimated using the exact radiation pattern of the quasi-BIC.